\documentclass[12pt,preprint]{aastex}

\accepted{}
\journalid{}{}
\articleid{}{}

\lefthead{Mirabal, Paerels \& Halpern}
\righthead{High-Resolution Spectroscopy of GRB 020405}

\received{2002 September 10}
\begin{document}

\def\etal{{\it et al.}}
\def\eg{{\it e.g.,}}
\def\ie{{\it i.e.,}}
\def\vs{{\it vs.}}
\def\etc{{\it etc.}}
\def\kms{km~s$^{-1}$}
\def\Msol{M$_\odot$}
\def\lsim{\mathrel{\lower .85ex\hbox{\rlap{$\sim$}\raise
.95ex\hbox{$<$} }}}
\def\gsim{\mathrel{\lower .80ex\hbox{\rlap{$\sim$}\raise
.90ex\hbox{$>$} }}}
\newbox\grsign \setbox\grsign=\hbox{$>$}
\newdimen\grdimen \grdimen=\ht\grsign
\newbox\laxbox \newbox\gaxbox
\setbox\gaxbox=\hbox{\raise.5ex\hbox{$>$}\llap
     {\lower.5ex\hbox{$\sim$}}}\ht1=\grdimen\dp1=0pt
\setbox\laxbox=\hbox{\raise.5ex\hbox{$<$}\llap
     {\lower.5ex\hbox{$\sim$}}}\ht2=\grdimen\dp2=0pt
\def\gax{\mathrel{\copy\gaxbox}}
\def\lax{\mathrel{\copy\laxbox}}

\def\ie{i.e.}
\def\xray{GRB~020405}
\def\source{GRB~020405}
\def\cha{{\em Chandra\/}}
\def\asca{{\em ASCA\/}}
\def\ro{{\it ROSAT\/}}

\def\pz{\phantom{0}}
\def\lsim{\mathrel{\lower .85ex\hbox{\rlap{$\sim$}\raise
.95ex\hbox{$<$} }}}
\def\gsim{\mathrel{\lower .80ex\hbox{\rlap{$\sim$}\raise
.90ex\hbox{$>$} }}}

\title{
High Resolution Grating Spectroscopy of GRB 020405 with
{\it CHANDRA\/} LETGS}

\author{N. Mirabal\altaffilmark{1}, F.~Paerels\altaffilmark{1} and
J.~P.~Halpern\altaffilmark{1}}
\altaffiltext{1}{Columbia Astrophysics Laboratory, Columbia University,
  550 West 120th Street, New York, NY~10027}

\begin{abstract}
\rightskip 0pt \pretolerance=100 \noindent
   We present high-resolution X-ray spectroscopy of GRB 020405
obtained with the Low Energy Transmission Grating Spectrometer (LETGS)
on board the \cha\ {\em X-ray Observatory\/} 
starting 1.68 days after the burst.
The spectrum appears featureless, with
no evidence for emission lines,
absorption edges, or narrow radiative recombination continua. 
The continuum can be fitted by a power law of photon index 
$\Gamma = 1.72 \pm 0.21$
and temporal decay index $\alpha = 1.87 \pm 0.1$, with a marginally 
significant
excess column density of cold gas 
$N_{H}$=$(4.7 \pm 3.7) \times 10^{21}$ cm$^{-2}$ at the 
redshift of the host galaxy.  
The absence of iron
lines indicates that the density of nearby surrounding
material was unlikely to be very dense
($n \lax 5 \times 10^{12}$ cm$^{-3}$) at the time of the
\cha~observation. In
the case of recombination following photoionization in an
optically thin medium,
most ionic species would be completely stripped at lower gas densities
than this.  In the case of a power-law spectrum reflecting off a 
``cold'', opaque
medium of low density, negligible emission features would be produced.
Alternative to these possible explanations for the lack of emission
features,
any X-ray line emission taking place in a dense medium
in a ``nearby reprocessor'' scenario might have been overwhelmed
by the bright afterglow continuum.
Although the absence of discrete features
does not unambiguously test for a connection between GRB 020405 and
nucleosynthesis,
it emphasizes the need for high-resolution X-ray spectroscopy to
determine
the exact emission mechanism responsible for the reported
discrete lines in other GRB afterglows.

\end{abstract}

\keywords{gamma rays: bursts --- X-Rays: general}

\section{Introduction}
The nature of gamma-ray burst (GRB) progenitors remains a mystery
despite the localization of $\approx$ 40
X-ray afterglows\footnote{See http://www.aip.de/$\sim$jcg/grbgen.html}.
An increasing number of X-ray observations reporting
discrete spectral features in GRB afterglows
(Piro et al. 1999; Antonelli et al. 2000; Piro et al. 2000)
has motivated attempts to find an explanation
in terms of GRB progenitors.
Thus far the claimed features are
consistent with n = 2 $\rightarrow$ 1 transitions in all
possible charge states of Fe, i.e., from Fe K-shell fluorescence at 6.4
keV
to H-like Fe Ly $\alpha$
at 6.95 keV. More recently Reeves et al. (2002)  have reported
the detection of
H-like emission from multiple $\alpha$-elements including
Si and S in the spectrum of GRB 011211; however, the statistical
significance of those features
is debatable (Rutledge \& Sako 2002). The possible
presence of Fe and/or $\alpha$-elements in the X-ray spectra of GRB
afterglows
could provide valuable information
about the physical conditions in the vicinity of the progenitors.

Determining the
relevant X-ray reprocessing mechanism in GRBs will
involve the
physical properties and geometry of the surrounding medium
(see for instance Lazzati, Campana, \& Ghisellini 1999; Paerels et al. 2000).
Discrete spectral features are
expected from hot gas when electron collisions
cause excitations from the ground state.
Emission lines can also arise
via recombination if continuum photons (either from the burst itself or
its afterglow) create a highly photoionized plasma.
In certain cases
fluorescence may be important if ions retain sufficient numbers of bound
electrons.
An important diagnostic that might discriminate between the different
alternatives
is the appearance of radiative recombination continua (RRC)
produced by the return of unbound electrons
to the ground state.  RRC from a photoionized plasma
resemble narrow emission lines, and could be used to verify
that recombination is occurring.
A 10 ks grating observation of GRB 991216 (Piro et al. 2000)
showed a possible iron RRC
that could represent the first evidence of
a recombining plasma near a GRB.  Nevertheless, and despite these
efforts, the
X-ray mechanism responsible for discrete features
in GRBs is still uncertain due to limited
spectral resolution and sensitivity of the existing
observations (Paerels et al. 2000).
In this letter we present results of the
LETGS/ACIS-S spectroscopy of GRB 020405,
and we show how density and geometry regulate
the feasibility of detecting discrete lines in X-ray afterglows.

\section{X-ray Observation of GRB 020405}

GRB 020405 triggered the Third Interplanetary Network (IPN)
on UT 2002 April 5.0288 with
a 25-100 keV fluence $\approx$ $3 \times 10^{-5}$~erg~cm$^{-2}$ (Hurley
et al. 2002) making it at the time
the highest fluence above 25~keV for a localized
burst since  GRB 010629 (Ricker et al. 2001).
The localization of a bright optical
transient (Price, Schmidt, \& Axelrod 2002a) and a
relatively low foreground Galactic column density of $N_{\rm H}$ = 4.3
$\times$
$10^{20}$ cm$^{-2}$ along this line of sight
generated immediate interest in obtaining
high-resolution X-ray grating spectroscopy of its afterglow.

After a timely response by the \cha\ {\em X-ray Observatory\/}, the observation 
of GRB 020405 started on UT 2002
April 6.711 using the LETGS in conjunction with
the ACIS detector for a total usable exposure time
of 50.6 ks before the satellite went into Earth occultation
near perigee. At the time of
the final \cha\ trigger
we still lacked knowledge of the redshift,
therefore the observation was designed with a displacement along the
grating dispersion axis, Y-OFFSET=$2.\!^{\prime}45$. The offset
placed the 0th-order image on
the  front-illuminated ACIS-S2 chip (Figure 1) while the ACIS-S3 and
ACIS-S1
backside-illuminated chips, with higher quantum efficiency,
covered  higher spectral orders in the wavelength range
1.7 \AA~$\leq$~$\lambda$ $\leq$ 28.9
\AA~and 28.5 \AA~$\leq$  $\lambda$ $\leq$ 55.8 \AA~respectively.
A priori, this configuration includes redshifted H-like lines of
Ni, Co, Fe, Ca, Ar, S, Si, Al, Mg, Ne, and O
as well as a few L-edges and He-like
lines from a number of metals
assuming an average GRB redshift$^{2}$ $z \sim 1.3$.

\section{Temporal and Spectral Analysis}

The standard \cha\ data products were processed with the
destreak tool to correct a minor flaw in the readout of
the chips\footnote{http://asc.harvard.edu/ciao/threads/destreak/}.
Subsequently, the
instrument and mirror response were generated using the ``makearf'' and
``makermf'' tools. The final 0th and
1st-order spectra were then extracted using the proper
grating effective area function (ARF) and redistribution matrix file
(RMF)
for each setting. For the timing analysis we used counts
extracted from a circular aperture centered on the 0th-order image
(we obtain consistent results using the dispersed data). A light curve
of the LETGS events as a function of time
shows a decay of the X-ray afterglow of GRB
020405 (Figure 2). The flux can be characterized as a power law
$\propto t^{-\alpha}$ (where t stands for time after the initial burst
event) with $\alpha = 1.87 \pm 0.1$.

The spectral analysis was undertaken using the spectral fitting package
XSPEC.
We analyzed the
+1 and --1 orders individually as well as coadded.
The results obtained in each case are statistically consistent.
A fit of the
coadded 1st-order extracted spectrum is well represented
by an absorbed power law
with photon index $\Gamma$ = 1.74 $\pm$ 0.22 and column density
$N_{\rm H}$ = (1.94 $\pm$  1.14)$\times 10^{21}$ cm$^{-2}$
($\chi_{\nu}^{2}$ = 0.26). Rather than leaving $N_{\rm H}$ as a
free parameter, we also fitted a power-law model with Galactic
and redshifted absorption at the host redshift, $z=0.690$
(Masetti et al. 2002; Price et al. 2002b) by treating the Galactic
absorption along the line of sight,
$N_{\rm H}$ = 4.3$\times 10^{20}$ cm$^{-2}$, as a fixed parameter.
The best fit for such a model
has $\Gamma$ = 1.72 $\pm$ 0.21 with an absorbing column density at the
host $N_{\rm H}({\rm host}) = (4.7 \pm  3.7) \times 
10^{21}$cm$^{-2}$ ($\chi^{2}_{\nu}$ = 0.26).
Figure 3 shows the coadded 1st-order spectrum and power-law fit
including residuals.  The X-ray flux
$F_x$(0.2-10 keV) = (1.36 $\pm$ 0.25) $\times 10^{-12}$
ergs~s$^{-1}$~cm$^{-2}$, measured 1.71 days
after the burst, with a normalization
constant $A=2.41$ $\times 10^{-4}$
photons~keV$^{-1}$~cm$^{-2}$~s$^{-1}$ at 1~keV
is similar to the flux reported for the bright GRB 010222
at the same epoch (in't Zand et al. 2001).
Although the value of $N_{\rm H}$
is not highly constraining, it suggests that absorption at the host is
non-negligible.

The X-ray flux, in comparison with optical photometry obtained
around the same epoch, 1.7 days after the burst
(Bersier et al. 2002, Price et al 2002b),
can be used to derive a broad-band slope $\beta_{ox} = 0.74$.
Remarkably, this is identical to the X-ray spectral index itself,
$\beta_x = \Gamma - 1 = 0.72 \pm 0.21$.  However, a smooth
extrapolation through the $BVRI$ photometric points is {\it not\/}
indicated, since fits to the $BVRI$ spectrum yield slope
$\beta_o = 1.25$ (Price et al. 2002b) or $\beta_o = 1.43$
(Bersier et al. 2002), the later even after correcting for
Galactic extinction.  Thus, the optical slope is steeper than the X-ray,
seeming to require somewhere a concave upward inflection in the spectrum,
which is inconsistent with basic synchrotron afterglow theory.
This kind of discrepancy is common in afterglow spectra
(e.g., Halpern et al. 1998), and is most simply understood
as requiring additional dereddening of the optical spectrum
to account for local extinction in the host.  In this case, the
marginally significant $N_{\rm H}({\rm host})$ derived from
the fit to the X-ray spectrum, if accompanied by typical
amounts of optical extinction, would be enough to restore a
concave downward shape to the broad-band spectrum.

Alternatively, this broad-band spectrum
can be described as having an X-ray excess, which would
require a mechanism in addition to synchrotron emission to
be operating.  For example, an X-ray contribution from 
inverse Compton scattering may be indicated,
as was deduced by modelling of the broad-band afterglow
of GRB 000926 by Harrison et al. (2001).

Of the simple synchrotron models
involving expansion into a uniform-density medium (Sari,
Piran, \& Halpern 1999), the observed combinations of spectral
slope and temporal decay in the optical and X-ray 
are most consistent with the evolution of a jet-like afterglow
after both the cooling break and the jet break have passed.
However, the required electron energy power-law slope $p$
would be slightly less than 2, which violates
the assumptions of the model from which the relations
$\alpha = p$ and $\beta = p/2$ were derived.  A different
environment in which the medium has a stellar-wind density profile
$n \propto r^{-2}$ should be considered, because its predictions
allow a more reasonable value of $p$.  In the adiabatic case
where the cooling frequency has not yet passed, $\alpha = (3p-1)/4 =
(3\beta + 1)/2$ (Chevalier \& Li 2000).  For example, an acceptable fit
in the wind scenario has $p = 2.7$, $\alpha = 1.78$,
and $\beta = 0.85$.  A wind model should be of interest in the
context of evidence for a massive progenitor star from
the possible detection of a supernova in the late-time
decay curve (Price et al. 2002b).

The 1st-order spectrum for the
whole observation shows no evidence for
discrete emission features,
absorption edges or narrow RRC signatures across the spectrum.
The edge seen around 5.62 \AA~corresponds to
an instrumental absorption edge in gold material used in the LETGS.
In addition to searching for features in the summed observation,
the data were divided into equal time intervals of various durations
($\Delta T$= 5.1 ks, 10.1 ks, 16.87 ks) to test for transient features
that
might have been present for part of the observation, as was reported in
the case
of GRB 970508 (Piro et al. 1999). The search for
transient features did not produce any detection
at the 3 $\sigma$ level.
We also looked for evidence of discrete features in the
0th-order spectrum.
Although there are no features evident, the detection limits are not as
sensitive as in the dispersed spectra because of
the reduced spectral resolution and efficiency of the ACIS-S2 chip.

In order to quantify the absence of discrete features,
we make a simplified assumption and adopt
a fixed discrete line width $\sigma_{\rm rest}$ = 0.46 keV in the GRB
rest frame ($\sigma_{\rm obs}$ = 0.27 keV at $z = 0.690$).
This value is borrowed from the Fe line measurement in the
GRB 991216 spectrum as reported by Piro et al. 2000.
We then proceeded to fit
Gaussian line profiles at their redshifted energies for multiple
transitions and used the power-law continuum level
to derive the line upper limits listed in Table 1. Finally,
we determined upper limits for the rest-frame equivalent widths
at the redshifted location of the transitions listed in Table 1.

\section{X-ray Line Emission Mechanisms}

The main objective of this grating spectroscopy was to acquire
a sensitive soft X-ray spectrum to derive constraints on material
in the proximity of the GRB. This is in fact the brightest case where
X-ray spectroscopy has been obtained for a GRB in which a SN-like
signature in the late-time decay curve (Price et al. 2002b) has been reported.
A significant
detection of Fe peak elements and particularly
medium-$Z$ $\alpha$ elements (C through Si), where the highest spectral
sensitivity for this choice of instrument is achieved, could have
helped establish the nature of GRB progenitors.
Unfortunately, the absence of discrete features in the spectrum of GRB
020405 does not provide a direct link between GRB 020405
and nucleosynthesis. Nevertheless we use the observed X-ray spectrum
to place tentative constraints on the relevant
emission mechanisms.

\subsection{Collisional Excitation}

There has been a resurgence of interest in the possible
role of collisional excitation
as a line emission mechanism in GRBs.
The reason is mainly the MEKAL plasma model
fit to the reported spectral emission lines
from GRB 011211 (Reeves et al.
2002).  A brief review of the
thermal history of GRB 020405 is needed to evaluate the
importance of collisional excitation to its X-ray afterglow emission.
During the first 60 seconds, the
burst bathes the progenitor gas with photons of average energy
$E \gax$ 0.1 MeV (Price et al. 2002b).
For the prompt spectrum and peak energy
of the burst $E_{\rm p}$ (Band et al. 1993),
the photon temperature averages $T_{\rm prompt} \sim 10^{9}$ K. However,
we note that the Compton time scale required to equilibrate is much
longer than the duration of the burst unless the electron number
density $n \gax 10^{12}$ cm$^{-3}$ (Paerels et al. 2000). In other
words,
unless the medium is very dense, the Compton temperature of the prompt
emission will not
be reached. Current numerical models of proposed
progenitor collapse have considered central
densities of the required order (MacFadyen, Woosley, \& Heger 2001).
Nonetheless, even if occurring at early times, such a transient
effect will
probably be short lived because the plasma cools as
the afterglow begins. In fact, the plasma temperature will settle to a
new
value corresponding to
the balance between Compton heating and cooling and determined by the
afterglow spectral parameters. Consequently, in the nonrelativistic
limit,
the temperature of the emitting material should not exceed the afterglow
Compton temperature $T_{\rm C}$ given by Ross, Fabian, \& Young (1999)
and expressed as
\begin{equation}
  4kT_{\rm C}\int_{E_1}^{E_2} u_E\,dE =
  \int_{E_1}^{E_2} u_E\left(E-{21E^2\over 5m_{\rm e}c^2}\right)dE,
\end{equation}
where $u_E\propto E^{1-\Gamma}$, $E_{1}$ = 0.1 keV, and
$E_{2}$ = 100.0 keV gives
$T_{\rm C} \sim$ $4 \times$ $10^{7}K$ for $\Gamma = 1.72$.

Additional losses due to bremsstrahlung and atomic transitions continue
to
cool the plasma below $T_{\rm C}$
as the cooling time scale becomes comparable to the duration of the
afterglow.
The absence of statistically
significant emission in the X-ray spectrum implies that collisional
excitation
does not play a significant role in driving line emission
by the time of the X-ray observation.
In short, collisional excitation by
afterglow photons might be important only when there is a
sufficiently enough dense medium ($n \gax 10^{10}$ cm$^{3}$)
for Compton interactions to equilibrate, and if an additional
heating mechanism is present that can maintain the gas around
temperatures $T \gax 10^{8} K$, as suggested by Paerels et al. (2000).

\subsection{Recombination \& Fluorescence}

As the ionizing afterglow continuum encounters the external medium,
discrete
emission might also be excited by recombination or fluorescence.
Depending on
the optical depth and geometry of the medium, we can refer to this
mechanism as
either transmission or reflection
(Vietri et al. 2001).  In transmission,
an optically thin medium ($\tau < 1$) might
produce photon cascades following recombination. In reflection,
emission can be produced by the afterglow spectrum reflecting
off an optically thick slab ($\tau \gg$ 1). Although the
line equivalent widths of order a few keV reported in prior
observations favor reflection over transmission, we
will consider both possibilities here.
The structure of recombination in either case  will be
determined by the ionization parameter $\xi \equiv
{L_{x}/nR^{2}}$ (Tarter, Tucker, \& Salpeter 1969)
where $L_{x}$ is the ionizing luminosity,
$n$ is the number density and $R$ the distance from the ionizing source.

To start, we derive the constraint on density for recombination
taking place in an optically thin, homogeneous medium. Nominally, since
the
\cha~observation ended 2.27 days after the GRB
any emitting material must have been located within a radius $R$
of the GRB where
\begin{equation}
R\, = \, {ct_{\rm obs} \over (1+z)} \, {1\over (1-\cos\theta)}\, \approx
\
{3.5 \times 10^{15}} {1\over (1-\cos\theta)}\ {\rm cm},
\end{equation}
where  $\theta$ is the angle between the observer
and the emitting material as seen from the GRB. Furthermore the
light-distance
traveled in the rest frame
in the duration of the X-ray observation is $\Delta R = 8.98 \times
10^{14}$ cm.

Next we adopt a time-dependent X-ray luminosity

\begin{equation}
L_{x}(0.2-10\ {\rm keV}) \approx 3.3 \times
10^{45}{{\Omega}\over{4\pi}} \left (t \over 1.71\, {\rm
days}\right)^{-1.87}
{\rm ergs\ s^{-1}}
\end{equation}
with energy index $\beta_x=0.72$ where $\Omega/4\pi$ corresponds
to the jet solid angle. Price et al. (2002b) estimated
a jet opening angle of $(5.\!^{\circ}83 \pm 0.\!^{\circ}69)\,n^{1/8}$
for GRB 020405
where $n$ is the particle density in the nearby medium. However, since
the
solid angle of the X-ray continuum is poorly constrained at the time of our
observation, which may overlap the ``jet break,''
we leave $\Omega$ as a free parameter in the following discussion.
At $z=0.690$, GRB 020405 has a luminosity
distance $d_{L}$ = 4505 Mpc, assuming a
$H_0 = 65$ km~s$^{-1}$~Mpc$^{-1}$, $\Omega_{\rm m} = 0.3$,
$\Omega_\Lambda = 0.7$ cosmology. Hence
\begin{equation}
\xi
\approx {{3.3 \times
10^{2}}\over{n_{11}r_{16}^{2}}}{{\Omega}\over{4\pi}} \left(t \over
1.71\, {\rm days}\right)^{-1.87}
\end{equation}
where
$r_{16}$ is the distance from the ionizing source in units of $10^{16}$
cm
and $n_{11}$ the number density in units of $10^{11}$ cm$^{-3}$.

As noted above, the state of the gas, including the ionization fraction
$f$
for each species, is largely determined by the ionization parameter
$\xi$.
In order to estimate $f$ more accurately as a function of $\xi$,
we ran the XSTAR code (Kallman \& Bautista 2001).
XSTAR produces ionization fractions for
H, He, C, N, O, Ne, Mg, Si, S,
Ar, Ca, Fe, and Ni using the appropriate photoionization and
recombination
rates. For our model we adopted an input
with power-law energy index $\beta_x$=0.72 and a homogeneous medium with
solar chemical composition within the radius of emission $R$. The output
of XSTAR shows that the ionization fraction of bound Fe$^{+25}$ starts to be
significant near
$\xi \sim 10^{5}$, reaching a maximum at $\xi \sim 10^{4}$.
Other species require lower ionization parameters for significant
ionization
stages with bound electrons to be present.

In order to constrain the density of the medium,
we express the
recombination luminosity of any line $L_{\rm line}$ as
\begin{equation}
{L_{\rm line} \over L_x} \simeq 4 \pi n \Delta R
\alpha(Z,T) f A_Z {E_{\rm line}\over
\xi}
\end{equation}
where $A_Z$ is the element abundance, $E_{\rm line}$ is the line
energy,
$f$ the ionization fraction,
$\alpha(Z,T)$ the recombination coefficient (Seaton 1959).
Assuming $\xi \lax 10^{5}$,
an ionization fraction of order 0.1 (possibly overestimated),
and standard solar abundances, the upper limit to
the Fe K$\alpha$ line luminosity listed in Table 1
translates into an upper limit in density
\begin{equation}
n \lax 5 \times 10^{12} {{4\pi}\over{\Omega}}\ {\rm cm}^{-3}
\end{equation}
This expression
shows that significant ionization fractions require
densities $n$ $\gax$ 5 $\times$ $10^{12}$ cm$^{-3}$ for
species to be recombining under conditions of ionization equilibrium.
For lower
densities, most species including Fe would be completely
stripped, which would account for the absence of discrete features
in the spectrum of GRB 020405. Moreover, the medium becomes optically 
thick near n $\gax$  2 $\times$ $10^{9}$ cm$^{-3}$.
The range of ionization
fraction maxima for Fe$^{+25}$ and other species is relatively narrow
and could
serve as an indicator of the recombination mechanism in transmission
as $\xi$ evolves with time due to the afterglow decay. Another
caveat in this analysis is our assumed standard solar abundances.
Supersolar
compositions (10--100 times solar)
would relax the restrictions on density by the appropriate
factor, as the line luminosity is linearly dependent on both abundance
and density.

The reflection model for an
afterglow spectrum off a ``cold'', optically thick medium
has been investigated numerically by a number of groups
(Ballantyne \& Ramirez-Ruiz 2001;
Lazzati, Ramirez-Ruiz, \& Rees 2002). In particular, the reflection
model has been put forth to explain the X-ray
spectrum of GRB 991216 (see for instance Vietri et al. 2001).
Using XSTAR once more
with the estimated power law spectrum, solar abundance 
and radius of emission $R$, we find that
reflection models appear most efficient for ionization parameters
$\xi \lax 10^{4}$. This result is consistent with previous
work (Ballantyne \& Ramirez-Ruiz 2001; Kallman, M\'esz\'aros, \& Rees
2001).
When line emission from Fe and other elements is absent, any geometry
with the optimal ionization parameter ($\xi \lax 10^{4}$)
implies a particle density of order
$n \lax 2.7 \times 10^{10}$ cm$^{-3}$ at the assumed emitting region
$R$. Lower column densities
increase the ionization parameter and the quoted models tend to
become near-perfect reflectors thus exhibiting negligible emission
features
(Ross, Fabian, \& Brandt 1996).
A near-perfect reflector is also consistent
with the absence of emission features in the spectrum
of GRB 020405.

Below the ionization stage Fe$^{+23}$, Fe K$\alpha$ fluorescence
emission might follow photoionization.
Although the fluorescent line luminosity
is typically independent of the ionization
parameter $\xi$, fluorescence is usually
important when $\xi \leq 10^{3}$ (Kallman, M\'esz\'aros, \& Rees 2001).
This
implies that fluorescence might only occur
if $n \gax 2.7 \times 10^{11}$ cm$^{-3}$ for
emission material within $R$. The non-detection of Fe K$\alpha$ at
6.4 keV appears to rule out significant
fluorescence in a solar abundant medium.

\subsection{Afterglow Continuum: The Photon Curtain Effect}

Two general classes of models have been put forward to explain
the reported GRB X-ray features. In the first instance,
the ``distant reprocessor'' model assumes that the emitting material
in a optically thick shell or ring
is initially far enough from the afterglow front.
Possible conditions of this
scenario were discussed in \S 4.2 (See also
Weth et al. 2000; B\"ottcher 2000). On the
other hand, the ``nearby reprocessor'' model requires stellar-like
gas densities near the GRB site and an ionizing continuum provided
by an expanding bubble of magnetized plasma breaking through the stellar
envelope (M\'esz\'aros \& Rees 2001), by a pair screen scattering  
a substantial fraction of $\gamma$-rays backward (Kumar \& Narayan 2002), 
or by continuing emission from
the GRB source, e.g., a relativistic wind from a millisecond pulsar
(Rees \& M\'esz\'aros 2000).
One difficulty when considering the ``nearby reprocessor'' model
under the absence of significant line emission arises from our basic
ignorance of the continuing ionizing source close to the progenitor.
This is because the afterglow photons emitted by the jet at larger radii
render
disentangling the contribution close to the progenitor an almost infeasible
task
(Rees \& M\'esz\'aros 2000). This
effect is problematic for the detection of line emission as well
since a
``photon curtain'' of afterglow continuum could swamp any discrete
line emission. In GRB 020405, the bright
continuum could mask line emission
occurring well within a radius of emission ($\ll R$)~even for dense
media ($n \gg 10^{12}$ cm$^{-13}$).
In summary, a long-lasting, bright
afterglow continuum might hinder the detection
of faint discrete line
in the ``nearby reprocessor'' model unless favorable conditions that
might attenuate the afterglow continuum (\ie~off-axis explosions,
dark bursts?)
are in place at the time of the observation.

\section{Conclusions and Future Observations}

Our analysis of various discrete X-ray emission models show that the
pure power-law spectrum of GRB 020405 is adequately explained
by a low-density medium with $n \lax 5 \times 10^{12}$ cm$^{-3}$, or
alternatively by a ``photon curtain'' effect produced by
the afterglow continuum in a dense ``nearby reprocessor''
scenario.
In the future, as the rapidity of localization improves, high-resolution
spectroscopy
could play a decisive role in resolving the emission mechanism (Paerels
et
al. 2000).

Figure 4 shows a simulated Gaussian
line profile for Fe K$\alpha$ with line width and flux identical to
those reported for GRB 991216 (Piro et al. 2000),
added to the observed power-law spectrum
of GRB 020405. A Fe K$\alpha$ line
with similar properties would have been clearly detected
in our LETGS spectrum.
But quite possibly the major accomplishment of the \cha~LETGS in grating
spectroscopy could come in the detection of
medium-$Z$ $\alpha$ elements, $Z < 14$, where its higher effective area
and
sensitivity are superior to other instruments on board \cha.
Continuous high-resolution spectroscopy might also provide an avenue to
detect a change in the emission
mechanism as a function of time \ie, from recombination to fluorescence.
In summary, our results confirm the need for
further high-resolution observations to improve the significance of
previous
claims of detections for discrete features in GRB afterglows.

\acknowledgments{ }
We are grateful to all the members of the Chandra team, especially Ed
Kellogg,
for timely planning, implementation and processing of this observation
under
less than ideal circumstances. We also thank  Kaya Mori and
Maurice Leutenegger for useful conversations. Finally
we acknowledge Paul Price and the Caltech GRB team for the
prompt localization of the optical counterpart.

\clearpage

\begin{deluxetable}{cccccr}
\tablenum{1}
\tablecolumns{5}
\tablewidth{0pc}
\tablecaption{Line Emission Upper Limits }
\tablehead
{
Ion              & $E_{\rm line}({\rm keV})$ &$L_{\rm line}(\times
10^{44}
{\rm ergs\ s^{-1}}$) &  EW({\rm keV})\tablenotemark{a} & $\alpha(Z,T)$
\tablenotemark{b}& \omit \hfil $A_Z$ \hfil}
\startdata
Ni XXVIII & 8.00 &  $\leq$ 5.0 & $\leq$ 2.33 & $\geq$ 3.42 $\times$
$10^{-12}$ & 1.78  $\times
$ $10^{-6}$\\
Co XXVII & 7.40 &  $\leq$ 3.5 & $\leq$ 1.63 & $\geq$ 3.11 $\times$
$10^{-12}$ & 8.60  $\times$
 $10^{-8}$\\
Fe XXVI & 6.97 & $\leq 2.5$ & $\leq$ 1.12 & $\geq 2.82 \times 10^{-12}$
& $4.68 \times 10^{-5}$\\
Fe XXV  & 6.70 & $\leq 2.3$ & $\leq$ 1.00 & & $4.68 \times 10^{-5}$\\
Fe XXIV & 6.40 & $\leq 2.4$ & $\leq$ 1.03 & & $4.68 \times 10^{-5}$\\
Ca XX & 4.11 &  $\leq$ 3.9 & $\leq$ 1.42 & $\geq$ 1.42 $\times$
$10^{-12}$ & 2.29  $\times$ $1
0^{-6}$\\
Ca XIX & 3.90 &  $\leq$ 4.5 & $\leq$ 1.60  & & 2.29  $\times$
$10^{-6}$\\
Ar XVIII & 3.32 &  $\leq$ 4.7 & $\leq$ 1.58 & $\geq$ 1.07 $\times$
$10^{-12}$ & 3.63  $\times$
 $10^{-6}$\\
Ar XVII & 3.14 &  $\leq$ 3.7 & $\leq$ 1.22 & & 3.63  $\times$
$10^{-6}$\\
S XVI & 2.62 &  $\leq$ 3.1 & $\leq$ 0.95 & $\geq$ 7.77 $\times$
$10^{-13}$ & 1.62  $\times$ $1
0^{-5}$\\
S XV  & 2.46 &  $\leq$ 5.2 & $\leq$ 1.56 & & 1.62  $\times$ $10^{-5}$\\
Si XIV & 2.01 &  $\leq$ 4.7 & $\leq$ 1.42 & $\geq$ 5.39 $\times$
$10^{-13}$ & 3.55  $\times$ $
10^{-5}$\\
Al XIII & 1.73 &  $\leq$ 5.6 & $\leq$ 1.47 & $\geq$ 4.39$\times$
$10^{-13}$ & 2.95  $\times$ $
10^{-6}$\\
Mg XII & 1.47 &  $\leq$ 6.9 & $\leq$ 1.71 & $\geq$ 3.51 $\times$
$10^{-13}$ & 3.80  $\times$ $
10^{-5}$\\
Mg XI & 1.35 &  $\leq$ 7.9 & $\leq$ 1.88 & & 3.80  $\times$ $10^{-5}$\\
Na XI & 1.24 &  $\leq$ 4.4 & $\leq$ 1.02 & $\geq$ 2.75 $\times$
$10^{-13}$ & 2.14 $\times$ $10
^{-6}$\\
Ne X & 1.02 &  $\leq$ 5.5 & $\lax$ 1.15 & $\geq$ 2.09 $\times$
$10^{-13}$ & 1.23  $\times$ $10
^{-4}$\\
Ne IX & 0.92 &  $\leq$ 6.4 & $\lax$ 1.24 & & 1.23  $\times$ $10^{-4}$\\
O VIII & 0.65 &  $\leq$ 6.1 & $\lax$ 0.93 & $\geq$ 1.10 $\times$
$10^{-13}$ & 8.51  $\times$ $
10^{-4}$\\
O VII & 0.57 &  $\leq$ 6.0 & $\lax$ 0.82 & & 8.51  $\times$ $10^{-4}$\\
N VII & 0.50 &  $\leq$ 7.1 & $\lax$ 0.87 & $\geq$ 7.45 $\times$
$10^{-14}$ & 1.12  $\times$ $1
0^{-4}$\\
C VI & 0.37 &  $\leq$ 6.8 & $\lax$ 0.65 & $\geq$ 4.75 $\times$
$10^{-14}$ & 3.63  $\times$ $1
0^{-4}$\\
\enddata
\tablenotetext{a}{Rest-frame equivalent width measurement based on a
fixed line width model and continuum level derived from the power-law fit.}
\tablenotetext{b}{Indicates lower limit of recombination
coefficient assuming T $\lax 4 \times 10^{7}$ K at the time of the observation.}
\end{deluxetable}
\clearpage

\begin{figure}[tbp] \label{lris} \figurenum{1}
\begin{center}
\epsscale{.70}
\plotone{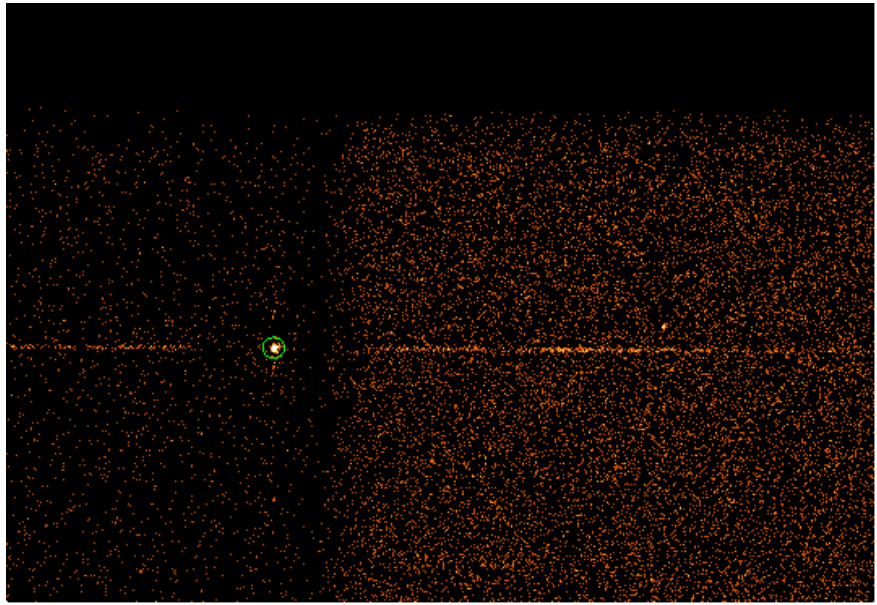}
\caption{Positioning of the 0th-order image on the front-illuminated S2 chip
and dispersed 1st-order spectra on the S2 and back-illuminated S3 chips
in the LETGS/ACIS-S configuration.}
\end{center}
\end{figure}
\clearpage

\begin{figure}[tbp] \label{esi} \figurenum{2}
\begin{center}
\epsscale{.70}
\plotone{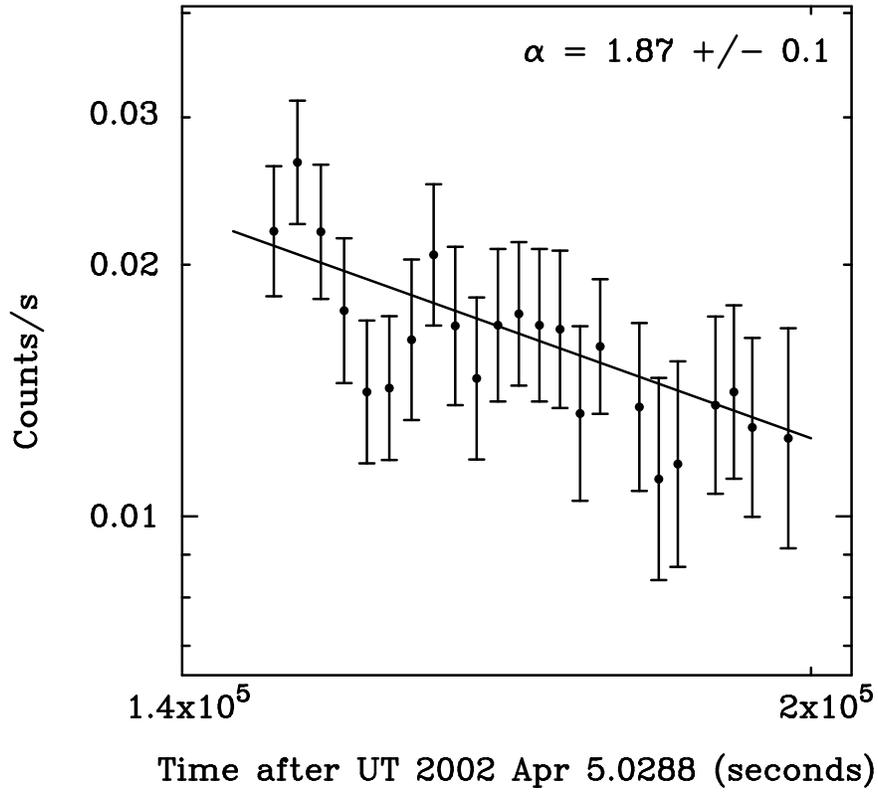}
\caption{X-ray light curve of GRB 020405 from the 0th-order LETGS image. 
The line shows the best-fitted power-law function describing the decay.}
\end{center}
\end{figure}
\clearpage

\begin{figure}[tbp] \label{optcont} \figurenum{3}
\begin{center}
\epsscale{.60}
\plotone{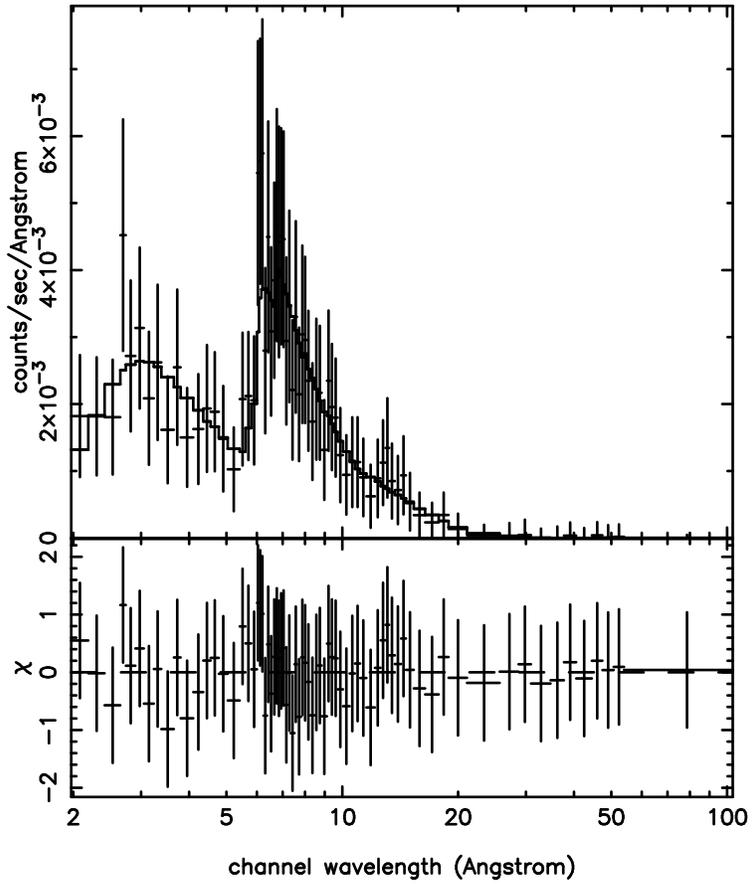}
\caption{The LETGS/ACIS-S 1st-order coadded spectrum for the whole
observation and the best-fitted absorbed power law model.
}
\end{center}
\end{figure}
\clearpage

\begin{figure}[tbp] \label{multilam} \figurenum{4}
\begin{center}
\plotone{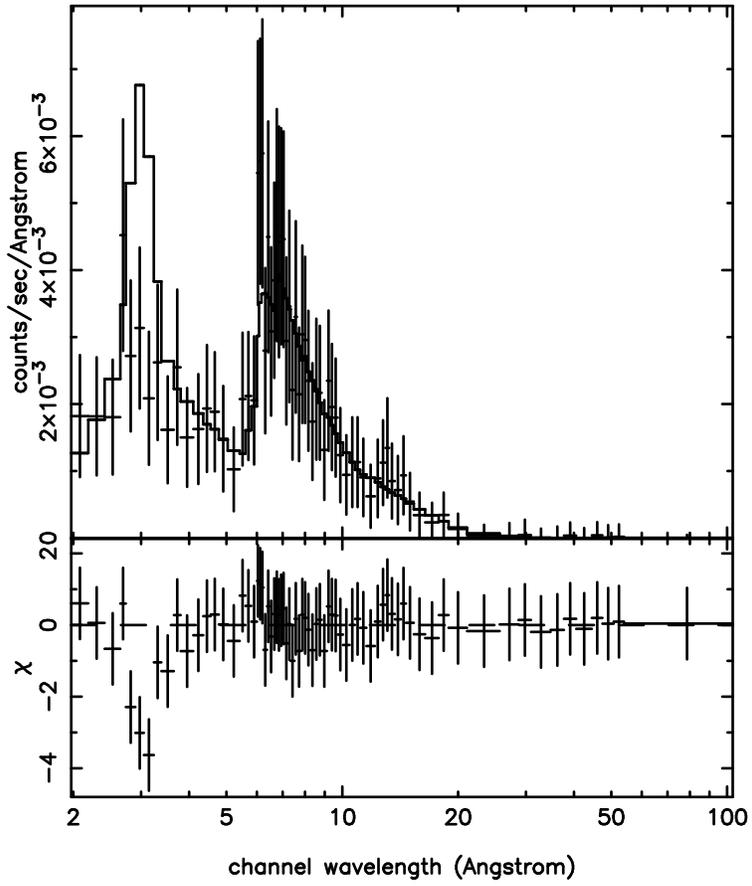}
\caption{Simulated Fe K$\alpha$ line model overlaid on the observed spectrum
of GRB 020405. Such a feature is clearly ruled out by the data.}
\end{center}
\end{figure}
\clearpage

\end{document}